\documentclass[11pt]{article}

\usepackage{amsmath}
\usepackage{amsfonts}
\usepackage{amssymb}

\def\be{\begin{eqnarray}}
\def\ee{\end{eqnarray}}
\def\nn{\nonumber}

\def\p{\partial}
\def\tr{{\rm tr}\,}

\hoffset= - 0.7in         

\title{{\bf Integrability of Hurwitz Partition Functions.
I.~Summary}
\vspace{.2cm}}
\author{{\bf A.Alexandrov}\footnote{ {\small {\it
LPTENS,  Paris , France}; {\it IPhT, Gif-sur-Yvette, France} and {\it ITEP, Moscow, Russia}};
}, {\bf A.Mironov}\footnote{ {\small {\it
Lebedev Physics Institute} and {\it ITEP, Moscow, Russia}};
mironov@itep.ru; mironov@lpi.ru}, {\bf A.Morozov}\thanks{{\small
{\it ITEP, Moscow, Russia}}; morozov@itep.ru} \ and {\bf
S.Natanzon}\thanks{{\small {\it ITEP, Moscow, Russia}; {\it
Department of Mathematics, Higher School of Economics, Moscow, Russia} and
{\it A.N.Belozersky Institute, Moscow State University, Moscow, Russia};
natanzons@mail.ru}}\date{ }}

\begin{document}

\maketitle

\vspace{-6.0cm}

\begin{center}
\hfill FIAN/TD-02/11\\
\hfill ITEP/TH-06/11\\
\hfill LPT ENS-11/11\\
\end{center}

\vspace{3.5cm}

\begin{abstract}
Partition functions often become $\tau$-functions of integrable hierarchies, if they are
considered dependent on infinite sets of parameters called time variables.
The Hurwitz partition functions
$Z = \sum_R d_R^{2-k}\chi_R(t^{(1)})...\chi_R(t^{(k)})\exp( \sum_n \xi_nC_R(n) )$
depend on two types of such time variables, $t$ and $\xi$.
KP/Toda integrability in $t$ requires that $k\leq 2$ and also that $C_R(n)$
are selected in a rather special way,
in particular the naive cut-and-join operators are not allowed for $n>2$.
Integrability in $\xi$ further restricts the choice of $C_R(n)$,
forbidding, for example, the free cumulants.
It also requires that $k\leq 1$.
The quasiclassical integrability (the WDVV equations) is naturally present
in $\xi$ variables, but also requires a careful definition of
the generating function.
\end{abstract}

\paragraph{1. Hurwitz numbers and Frobenius formula.}
Hurwitz numbers \cite{HuNu}
count the ramified coverings of a Riemann surface.
Their calculation is obviously important in string theory
and from time to time it attracts certain attention.
The current interest is due to a possibility of expressing the
Hurwitz numbers through group characters,
what implies close relation to matrix models,
integrable systems, Virasoro constraints,
AGT relation and other basic
chapters of modern theory.
This expression is given by the Frobenius formula \cite{Frof}
\be
{\rm Cover}_n(\Delta_1,\ldots,\Delta_k)
= \sum_R d^2_R\varphi_R(\Delta_1)\ldots\varphi_R(\Delta_k)
\delta_{|R|,n}
\label{Frof}
\ee
for the number of $n$-sheet coverings of a
Riemann sphere with $k$ ramification points of given types.
The type of ramification at a point $i$
characterizes the way in which the sheets are glued,
and is labeled by a Young diagram (integer partition of $n$)
$\Delta_i$ of weight $|\Delta_i|=n$.
The sum in (\ref{Frof}) goes over all the Young diagrams
$R = \{r_1\geq r_2\geq\ldots\}$ of the same weight
$|R| = r_1+r_2+\ldots = n$, and
$\varphi_R(\Delta)$ are symmetric group characters,
to appear in eq.(\ref{varphidef}) below.

For $k=1$ one can extend the sum in (\ref{Frof})
to all $R$ \cite{AMMops}.
Remarkably, though the sum is now infinite,
this produces only a factor of $e=2.718\ldots$
\be\label{2}
\sum_R d^2_R\varphi_R(\Delta)\delta_{|R|,|\Delta|} =
\frac{1}{e}\sum_R d^2_R\varphi_R(\Delta)
\ee
This motivates the change of the definition (\ref{Frof}).
Instead of (\ref{Frof}) one can write
\be
{\rm Cover}(\Delta_1,\ldots,\Delta_k)
= \frac{1}{e}\sum_R d^2_R\varphi_R(\Delta_1)\ldots\varphi_R(\Delta_k)
\label{Frofg}
\ee
without the $\delta_{|R|,n}$ projector.
In this form the r.h.s. is defined for arbitrary $\Delta_i$
and the index $n$ can be omitted.
These generalized Hurwitz numbers \cite{MMN1}
are much more interesting, but for $k>1$ they
are different from (\ref{Frof}), even when all
the diagrams have the same sizes,
$|\Delta_1|=\ldots = |\Delta_k|$.

\paragraph{2. Hurwitz partition functions and two ways to introduce
time-variables.}
To put the problem into a string theory context it remains
to substitute particular Hurwitz numbers by generating functions.
This can be done in two ways.

First, one can keep $k$ fixed and sum over all types
of ramification at a given point.
The clever way to do this is to use the crucial property
of $\varphi_R(\Delta)$ as expansion coefficients of the
Schur functions ($GL(\infty)$ characters) $\chi_R(t)$:
\be
\chi_R(t) = \sum_\Delta d_R\varphi_R(\Delta) p\,(\Delta)
\delta_{|\Delta|,|R|}
\label{varphidef}
\ee
where
\be
\chi_R(t) = \det_{ij} S_{r_i-i+j}(t), \nn \\
\exp \left(\sum_k t_kz^k\right) = \sum_k z^kS_k(t)
\ee
$p_k = kt_k$ and for $\Delta = \{\delta_1\geq\delta_2\geq\ldots\}$
we define
$p\,(\Delta) = p_{\delta_1}p_{\delta_2}\ldots = p_1^{m_1}p_2^{m_2}\ldots$
and, for the future use,
$\widetilde p\,(\Delta) = p\,(\Delta)/z(\Delta)$,
where $z(\Delta) = \prod_k m_k! k^{m_k}$.\footnote{
This combinatorial coefficient that counts the order of the
automorphism group of the Young diagram, appears everywhere in
the  theory of symmetric functions and symmetric group $S(\infty)$.
In particular, the standardly normalized symmetric group characters
$\hat\chi_R(\Delta)$
$$\hat\chi_R(\Delta) = d_R z(\Delta)\varphi_R(\Delta)$$
These $\hat\chi_R(\Delta)$ are generated by the command
${\bf Chi(R, \Delta)}$ in MAPLE in the package
{\bf combinat}, and we use
the hat to distinguish them from the linear group characters,
i.e. the Schur functions $\chi_R(t)$)
differ by this factor from our $\varphi_R(\Delta)$.}

In (\ref{varphidef}) one can remove projector $\delta_{|\Delta|,|R|}$
from the sum.
Since \cite{MMN1}
\be\label{defch}
\varphi_R(\Delta) =
\varphi_R(\tilde\Delta,1^{|\Delta|-|\tilde\Delta|})
= \frac{ (|R|-|\tilde\Delta|)! }
{(|R|-|\Delta|)!(|\Delta|-|\tilde\Delta|)!}
\varphi_R(\tilde\Delta,1^{|R|-|\tilde\Delta|})
\ee
where $\tilde\Delta$ is the sub-diagram of $\Delta$ which
does not contain unit lines,
in particular, $\varphi_R(\Delta) = 0$ for $|R|<|\Delta|$,
one has
\be
\chi_R(t+1) =
\chi_R(t_k+\delta_{k,1}) = \sum_\Delta d_R\varphi_R(\Delta) p\,(\Delta)
\label{chiphi}
\ee
i.e. the projector is removed at the expense of shifting the first
time $t_1$, for which we introduce a condensed notation $t\rightarrow t+1$.
For example, in this formula one can put all $t_k=0$, then only the
term with $\Delta=\emptyset$ contributes, and one gets
$d_R=\chi_R(\delta_{k,1})$.

Coming back to the generating function, one can use
(\ref{varphidef}) to convert (\ref{Frof}) into
\be\label{tZO}
Z\{t^{(1)},\ldots, t^{(k)}|q\}
= \sum_n q^n \sum_{\Delta_1,\ldots,\Delta_k}
{\rm Cover}_n(\Delta_1,\ldots,\Delta_k)
p^{(1)}(\Delta_1)\ldots p^{(k)}(\Delta_k)
\delta_{|\Delta_1|,n}\ldots \delta_{|\Delta_k|,n} = \nn \\
= \sum_n q^n \sum_R d_R^{2-k}\chi_R(t^{(1)})\ldots\chi_R(t^{(k)})
\delta_{|R|,n}
\ee
Remarkably, the generating function of the {\it generalized}
Hurwitz numbers (\ref{Frofg}) is given by the same formula(!),
with projector $\delta_{|R|,n}$ removed and substituted by the factor of
$1/e$ \cite{AMMops}:
\be
\sum_{\Delta_1,\ldots,\Delta_k}
{\rm Cover}(\Delta_1,\ldots,\Delta_k)
p^{(1)}(\Delta_1)\ldots p^{(k)}(\Delta_k) = \nn \\
= \frac{1}{e}\sum_R d_R^{2-k}\chi_R(t^{(1)}+1)\ldots\chi_R(t^{(k)}+1)=
{1\over e}Z\{t^{(1)}+1,\ldots, t^{(k)}+1|q=1\}
\label{tZ}
\ee
Non-unit $q$ can be introduced into this sum by the substitution $p_k\to q^kp_k$,
which leads to the factor of $q^{|R|}$ in the sum. Note that neither
the constant shift of $t$-variables, $t\rightarrow t+1$
nor the normalization factor of
$1/e$ is essential for
integrability properties below, therefore, we do not concern them
in what follows.
We emphasize that though (\ref{Frofg}) is different from (\ref{Frof})
the generating functions in (\ref{tZO}) and (\ref{tZ})
is the same function, only the arguments are shifted.

The second way to make a partition function is to
exponentiate $\varphi_R(\Delta)$:
\be
{\cal Z}_{excessive}\{\xi\} = \sum_R d_R^2
\exp\left( \sum_{\Delta} \xi_\Delta \varphi_R(\Delta)\right)
\label{xiZ}
\ee
This, however, introduces an excessive set of time-variables $\xi$,
labeled by all Young diagrams $\Delta$.
In the matrix model case, this would correspond to exponentiating
all multi-trace operators with independent couplings,
$\int dM \exp \left(\sum_\Delta \xi_\Delta \prod_k (\tr M^k)^{m_k}\right)$
(and, from integrable point of view to non-Cartanian hierarchies \cite{mamorev}).
However, this is not a clever choice, leading to anything nice:
instead one should better consider just
$\int dM \exp\left(\sum_k \xi_k \tr M^k\right)$.
Similarly, instead of (\ref{xiZ}), one better consider
\be
{\cal Z}_{C(n)}\{\xi\} =
\sum_R d_R^2 \exp\left(\sum_n \xi_n C_R(n)\right)
\ee
where $C_R(n)$ is {\it some} linear combination of $\varphi_R(\Delta)$,
one for each $n$.
The question is, however, what combination to choose,
and this is the main subject of our consideration below.
For historical reasons, $C_R(n)$ are often called
(eigenvalues of) Casimir operators.

Finally, one can consider the mixed partition function, with
both $t$ and $\xi$ variables:
\be
{\cal Z}(t|\xi) = \frac{1}{e}\sum_R d_R^{2-k}
\chi_R(t^{(1)}+1)\ldots\chi_R(t^{(k)}+1)
\exp\left(\sum_n \xi_n C_R(n)\right)
\ee
In particular, $q$ in (\ref{tZ}) is just $q = e^{\xi_1}$,
and $C_R(1) = \varphi_R(1)$ is defined unambiguously
because there is just one Young diagram of weight one.

\paragraph{3. Integrability properties: a summary.}
A cleverly defined partition function should be a $\tau$-function
of some integrable hierarchy \cite{mamorev}.
Though this is not necessary, in quite many cases the hierarchies
are "Cartanian": belong to the Toda/KP family associated with
the Kac-Moody algebra $\widehat{U(1)}$
(for more general $\tau$-functions see \cite{GKLMM}).
This turns out to be possible also for Hurwitz partition functions,
but imposes certain restrictions.
What is true is the following set of statements:

\begin{itemize}

\item {\bf Quasiclassical integrability.}

Quasiclassical integrability (WDVV equations)
in $\xi$-variables is most natural for Hurwitz partition function,
because of its clear algebraic topological nature.
In fact, this is most difficult kind of integrability,
it actually appears when the set of $\xi$ is {\it excessive}
and when partition function is defined with the help of the
$*$-product. This is a separate story, to be considered in \cite{WDVVmmn}.

\item
{\bf The basic example of t-integrability.}

$Z\{t^{(1)},\ldots,t^{(k)}|q\}$ is KP $\tau$-function in $t^{(1)}$
\underline{only} for $k=1$ and $k=2$. \be Z(t,\bar t|q) = \sum_R q^{|R|}
\chi_R(t)\chi_R(\bar t)=\exp\left(\sum_k kq^kt_k\bar t_k \right) \ee
is a KP $\tau$-function w.r.t. the both sets of times, $t_k$ and
$\bar t_k$.

This is a particular case of a more general statement:
\be\label{*}
\tau(t)=\sum_R w_R\chi_R(t)
\ee
is a KP $\tau$-function provided $w_R$ satisfy the bilinear Plucker relations
\be
w_{22}w_0-w_{21}w_1+w_2w_{11}=0\\
w_{32}w_0 - w_{31}w_1 + w_3w_{11} = 0\nn \\
w_{221}w_0 - w_{211}w_1 + w_2w_{111} = 0\nn\\
\hbox{...}\nn
\ee
which possess a solution $w_R=\det_{ij} A_{i,r_j-j}$ with any matrix $A_{ij}$
such that $A_{ij}=0$ for $i<0$ or
$j<0$ (i.e. non-zero only in the positive quadrant). In particular,
$w_R=\chi_R(\bar t)=\det_{ij} S_{i+r_j-j} (\bar t)$ for the Schur functions,
$\sum_rz^rS_r=e^{\sum z^kt_k}$. Moreover, one can
restrict the sum over $R$ in (\ref{*}) to the Young diagrams with finite number of
lines, $l(R)$:
\be\label{KPr}
\tau_{{\cal N}}(t)=\sum_{R:\ l(R)\le{\cal N}}w_R\chi_R(t)
\ee
It is still a KP $\tau$-function. The parameter ${\cal N}$ plays role of an additional
time-variable, "zero-time".

\item
{\bf KP $\tau$-function w.r.t. $(t,\bar t)$-variables.}

$\xi$-deformation preserves the $t$-integrability, i.e.
\be\label{KP1}
{\cal Z}_{C(n)}\{t,\bar t|\xi\} =
\sum_R \chi_R(t)\chi_R(\bar t) \exp \left(\sum_n \xi_n C_R(n)\right)
\ee
is a KP $\tau$-function in $t,\bar t$ variables \cite{GKM2} \underline{only} if
$C_R(n)$ is of the form
\be
\sum_n \xi_n C_R(n) = \sum_{k,i} \zeta_k \Big( (r_i-i)^k - (-i)^k\Big)
\label{Ct}
\ee
with arbitrary $\zeta_k$.

This is a restrictive condition:
in particular, the choice $C_R(n) = \varphi_R(n)$ with
single line diagrams is {\it not} allowed
beyond $n=1,2$. Indeed,
\be
\varphi_R(1) = \sum_j r_j = |R|, \ \ \ \ \ \ \ \ \
\varphi_R[2] =\frac{1}{2}\sum_j
\Big( (r_i-i+1/2)^2 -
(-i+1/2)^2\Big)
\ee
perfectly fits (\ref{Ct}), but already
\be
\varphi_R[3] =
\frac{1}{3}\sum_j r_j(r_j^2-3jr_j+3j^2-3j+2)
- \sum_{i<j} r_ir_j
\ee
is not of the form (\ref{Ct}) due to the last "mixing" term.
Thus, $\sum_R \chi_R\bar\chi_R e^{\xi_1\varphi_R(1)+\xi_2\varphi_R(2)}$
is, but $\sum_R \chi_R\bar\chi_R e^{\xi_3\varphi_R(3)}$
is {\it not} a KP $\tau$-function in $t,\bar t$.

\newpage

\item
{\bf Forced Toda lattice $\tau$-function w.r.t. $(t,\bar t)$-variables.}

In fact, (\ref{KP1}) can be further promoted to a Toda-lattice
$\tau$-function, which depends on additional zero-time
${\cal N}$, and, in addition to being a KP $\tau$-function
both in $t$ and $\bar t$ variables, satisfies an
extra equation
\be\label{1Todaeq}
\tau_{{\cal N}}{\partial^2\tau_{{\cal N}}\over\partial t_1\partial\bar t_1}-
{\partial\tau_{{\cal N}}\over\partial t_1}{\partial\tau_{{\cal N}}\over\partial\bar t_1}
=\tau_{{\cal N}+1}\tau_{{\cal N}-1}
\ee
However, for (\ref{KP1}) to satisfy this equation,
the $\zeta$-variables should depend on ${\cal N}$ in
a rather peculiar way.
Instead one can say that the Casimir operators in (\ref{KP1})
should be substituted by their peculiar ${\cal N}$-dependent
combinations:
\be\label{Toda1}
{\cal Z}_{C(n)}\{t,\bar t,{\cal N}|\xi\} =
e^{{\cal Q}_{{\cal N}}}\sum_{R:\ l(R)\le{\cal N}} \chi_R(t)\chi_R(\bar t)
\exp \left(\sum_n \xi_n C_{R;{\cal N}}(n)\right)
\ee
is a Toda-lattice $\tau$-function \underline{only} if
\be
\sum_n \xi_n C_{R;{\cal N}}(n) = \sum_{k,i}
\zeta_k \left((r_i+{\cal N}+\gamma-i)^k-({\cal N}+\gamma-i)^k\right)
=\sum_{n,k} \Big({{n}\atop{k}}\Big)\zeta_n C_{R}(k){\cal N}^{n-k}\nn\\
{\cal Q}_{{\cal N}}=\sum_{k}\sum_{i=1}^{{\cal N}}
\zeta_k ({\cal N}+\gamma-i)^k
\label{CtN}
\ee
where $\zeta_k$, $\gamma$ are arbitrary ${\cal N}$-independent constants.\\
Note that (\ref{Toda1}) can be rewritten in the form
\be
{\cal Z}_{C(n)}\{t,\bar t,{\cal N}|\xi\} =
\sum_{R:l(R)\le{\cal N}} \chi_R(t)\chi_R(\bar t)
\exp \left(\sum_n \xi_n \tilde C_{R;{\cal N}}(n)\right)
\ee
with
\be
\sum_n \xi_n \tilde C_{R;{\cal N}}(n) = \sum_{k}\sum_{i=1}^{{\cal N}}
\zeta_k (r_i+{\cal N}+\gamma-i)^k
\ee
where the sum over $i$ is now terminated at $i={\cal N}$
not automatically, but "by hands".\\
In fact, this is a $\tau$-function of \underline{forced}
Toda-lattice hierarchy \cite{KMMOZ,versus},
i.e. $\tau_0=1$ and $\tau_{{\cal N}}=0$ for ${\cal N}<0$.

\item
{\bf Toda lattice $\tau$-function w.r.t. $(t,\bar t)$-variables.}

One can lift this restriction by shifting ${\cal N}$ with constant ${\cal M}$
and then taking the limit ${\cal N},{\cal M}\to \infty$
in such a way that the new (shifted) zero-time
$N$ remains finite: ${\cal N}={\cal M}+N$.
With this procedure one is led to the $\tau$-function of the generic (unforced) Toda
lattice $\tau$-function:
\be
\tau_N\{t,\bar t|\xi\}=
e^{Q_N}\sum_{R} \chi_R(t)\chi_R(\bar t)
\exp \left(\sum_n \xi_n C_{R;N}(n)\right)
\ee
where the sum is now over {\it all} diagrams $R$,
independently of $N$, and
\be\label{QC}
\sum_n \xi_n C_{R}(n) = \sum_{k,i}
\zeta_k \left((r_i+N+\gamma-i)^k-(N+\gamma-i)^k\right)=\sum_{n,k}\Big({{n}\atop{k}}\Big)
\zeta_n C_{R}(k)N^{n-k}\nn\\
Q_{N}=\sum_{k}\sum_{i=1}^{N}
\zeta_k (N+\gamma-i)^k
\ee
A restriction of such a Toda lattice $\tau$-function
to just two non-vanishing $\xi_n$ and $\gamma=1/2$ appeared in
\cite{O}:
\be
\tau_N\{t,\bar t|\xi\}=
e^{\xi_1{N^2\over 2}+\xi_2{N(4N^2-1)\over 12}}
\underbrace{\sum_{R} \chi_R(t)\chi_R(\bar t)
e^{(\xi_1+2N\xi_2) C_{R}(1)+\xi_2 C_{R}(2)}}_{\tilde\tau(t,\bar t|q,\beta)}
\ee
Rescaled $\tau$-function $\tilde\tau(t,\bar t|q,\beta)$
with $q=e^{\xi_1+2N\xi_2}$ and $\beta=\xi_2/2$ satisfies the equation (see \cite[eq.(10)]{O})
\be
\tilde\tau(t,\bar t|q,\beta){\partial^2 \tilde\tau(t,\bar t|q,\beta)
\over\partial t_1\partial\bar t_1}-
{\tilde\tau(t,\bar t|q,\beta)\over\partial t_1}{\tilde\tau(t,\bar t|q,\beta)
\over\partial\bar t_1}
=q
\tilde\tau(t,\bar t|e^\beta q,\beta)\tilde\tau(t,\bar t|e^{-\beta}q,\beta)
\ee
which is a slight modification of (\ref{1Todaeq}), taking into account that
$e^{Q_{N+1}+Q_{N-1}-2Q_N}=q$.

\item
{\bf An example of $\xi$-integrability.}

Integrability in $\xi$-variables is even more restrictive:
\be
{\cal Z}_{C_n}\{\xi\} = \sum_R d_R^2
\exp\left(\sum_n \xi_n C_R(n)\right)
\label{Todachain}
\ee
is a KP $\tau$-function in $\xi$-variables only if
\be
\sum_n \xi_n C_R(n) = \sum_{n,i} \xi_n \Big( (r_i-i+\gamma)^n - (-i+\gamma)^n\Big)
\label{Cxi}
\ee
with arbitrary $\gamma$,
i.e. one can not choose the function
$\zeta(\xi)$ in (\ref{Ct}) in an arbitrary way:
only a very restricted class of {\it linear}
triangular changes $\{\xi\}\rightarrow\{\zeta\}$
in (\ref{Ct}) is allowed.

In particular, the expressions naturally emerging in theory of Kerov polynomials \cite{Kerov}
\be\label{Kerov}
\sum_n\xi_n C_R(n) = \sum_{k,i} \xi_{k}
\frac{1}{k+1}\Big((r_i-i+1)^{k+1} - (r_i-i)^{k+1}
- (-i+1)^{k+1} + (-i)^{k+1}\Big)
\ee
do not provide a $\tau$-function in $\xi$-variables.

\item {\bf Toda chain $\tau$-function in $\xi$-variables.}

Again, with the sum restricted to the Young diagrams
with ${\cal N}$ lines, like in (\ref{KPr}) or (\ref{Toda1}), one can consider
instead of (\ref{Todachain}) the generating function
\be\label{Todaxi}
{\cal Z}_{C(n)}\{t,\bar t,{\cal N}|\xi\} =
e^{{\cal Q}_{{\cal N}}}\sum_{R:\ l(R)\le{\cal N}} d_R^2
\exp \left(\sum_n \xi_n C_{R;{\cal N}}(n)\right)
\ee
It is a Toda chain $\tau$-function
in the $\xi$-variables,
with the zero-time ${\cal N}$ and
\be
\sum_n \xi_n C_{R;{\cal N}}(n) = \sum_{k,i}
\xi_k \left((r_i+{\cal N}+\gamma-i)^k-({\cal N}+\gamma-i)^k\right)
=\sum_{n,k} \Big({{n}\atop{k}}\Big)\zeta_n C_{R}(k){\cal N}^{n-k}\nn\\
{\cal Q}_{{\cal N}}=\sum_{k}\sum_{i=1}^{{\cal N}}
\xi_k ({\cal N}+\gamma-i)^k
\ee
The difference with (\ref{CtN}) is that now at the r.h.s. there should be $\xi_k$'s
instead of arbitrary coefficients $\zeta_k$'s.

This is the forced Toda chain $\tau$-function which satisfies the equation
\be
\tau_{{\cal N}}{\partial^2\tau_{{\cal N}}\over\partial \xi_1^2}-
\left({\partial\tau_{{\cal N}}\over\partial \xi_1}\right)^2
=\tau_{{\cal N}+1}\tau_{{\cal N}-1}
\ee
One can again repeat the procedure of removing the forced condition in order to obtain
\be
\tau_N\{t,\bar t|\xi\}=
e^{Q_N}\sum_{R} d_R^2
\exp \left(\sum_n \xi_n C_{R;N}(n)\right)
\ee
with
\be
\sum_n \xi_n C_{R}(n) = \sum_{k,i}
\xi_k \left((r_i+N+\gamma-i)^k-(N+\gamma-i)^k\right)=\sum_{n,k}\Big({{n}\atop{k}}\Big)
\xi_n C_{R}(k)N^{n-k}\nn\\
Q_{N}=\sum_{k}\sum_{i=1}^{N}
\xi_k (N+\gamma-i)^k
\ee
and again the difference with (\ref{QC}) is that here all $\zeta_k$'s
at the r.h.s. are replaced with $\xi_k$'s.

\item
{\bf KP $\tau$-function w.r.t. $(\xi,t)$-variables.}

$\xi$-integrability is preserved if $d_R^2$ is substituted by $d_Rw_R$,
where $w_R$ is any solution to the Plucker relations,
in particular, $w_R$ can be a character:
\be\label{37}
{\cal Z}_{C_n}\{t|\xi\} = \sum_R d_R\chi_R(t)
\exp\left(\sum_n\xi_n C_R(n)\right)
\ee
is a KP $\tau$-function both in $t$ and in $\xi$-variables,
provided Casimirs are chosen to satisfy (\ref{Cxi}).\\
However, $\sum_R \chi_R(t)\chi_R(\bar t)\exp \left(
\sum_n \xi_n C_R(n)\right)$,
though still a KP $\tau$-function in $t$ and $\bar t$ variables,
is {\it not} a KP $\tau$-function in $\xi$.\\
Also there is no way to introduce a ${\cal N}$-dependence
into (\ref{37}) to make it a Toda-lattice $\tau$-function.
This is in accordance with the general fact, that
a Toda-chain $\tau$-function can be promoted into
a Toda-lattice $\tau$-function only in a trivial way:
so that it
depends on $t$ and $\bar t$ only through  differences
$t_k-\bar t_k$.

\end{itemize}

\paragraph{4. Technical approaches.}
Technical details behind the checks and proofs of all these statements
will be presented in a separate publication.
They depend heavily on the theory of integrable hierarchies, however, these
are relatively old results. A principle new piece is the interplay with the newer
chapters of Hurwitz theory.
They are based on the study of associative and commutative algebra
(actually isomorphic to Kerov algebra \cite{KI})
of cut-and-join operators $\hat W(\Delta)$,
which have linear group and symmetric group characters $\chi_R(t)$
and $\varphi_R(\Delta)$ as their common eigenvectors and eigenvalues
respectively \cite{MMN1}:
\be\label{dr}
\hat W(\Delta)\chi_R(t) = \varphi_R(\Delta)\chi_R(t)
\ee
These operators can be represented as differential operators
in $t$-variables, for example,
\be
\hat W[2] = \sum_{a,b\geq 1} \left((a+b)p_ap_b\frac{\p}{\p p_{a+b}}
+ abp_{a+b}\frac{\p^2}{\p p_a\p p_b}\right)
= \sum_{a,b\geq 1} \left(abt_at_b\frac{\p}{\p t_{a+b}}
+ (a+b)t_{a+b}\frac{\p^2}{\p t_a\p t_b}\right)
\ee
or, after the Miwa transform $p_k = kt_k = \tr X^k$,
as elements of the center of the universal enveloping of $GL(\infty)$:
\be
\hat W(\Delta) = \ :\widetilde{\hat D(\Delta)}:\ =
\frac{1}{z(\Delta)}:\prod_k (\tr \hat D^k)^{m_k}:
\ee
with $\hat D_{ij} = \sum_k X_{ik}\frac{\p}{\p X_{jk}}$,
for example,
\be
\hat W[2] = \frac{1}{2!}:\tr \hat D^2:\ = \frac{1}{2}
\sum_{i,j,k,l} X_{ik}X_{jl}\frac{\p}{\p X_{jk}}\frac{\p}{\p X_{il}}
\ee
(If the factor of $z(\Delta)$ was omitted from the normalization of $\hat W(\Delta)$,
then, the eigenvalues would be $\frac{\hat\chi_R(\Delta)}{d_R}$.)
The structure constants are the same for multiplication of the
$\hat W$-operators and of their eigenvalues:
\be
\hat W(\Delta_1)\hat W(\Delta_2) = \sum_{\Delta} C_{\Delta_1\Delta_2}^\Delta
\hat W(\Delta), \nn \\
\varphi_R(\Delta_1)\varphi_R(\Delta_2) =
\sum_{\Delta} C_{\Delta_1\Delta_2}^\Delta\varphi_R(\Delta)
\ \ \ \ \forall R
\ee
and they are vanishing outside the interval
$max(|\Delta_1|,|\Delta_2|)\leq |\Delta| \leq
|\Delta_1|+|\Delta_2|$.
This algebra has various sets $\{\hat C(n)\}$
of
{\it multiplicative} generators, with one $\hat C(n)$
at each level $|\Delta|=n$.
$C_R(n)$ are their eigenvalues,
\be
\hat C(n)\chi_R(t) = C_R(n)\chi_R(t)
\ee
An obvious choice is to take $\{\hat W[n]\}$ with
single line diagrams for such a set,
but, as explained in s.3, this is {\it not} the choice,
preserving {\it any} kind of integrability.
Actually, the $t$-integrability can be preserved if $\hat C(n)$
are chosen to be free cumulants (\ref{Kerov}), whose (non-linear)
relation to $\hat W[n]$ (represented by Kerov's polynomials) is known from \cite{Kerov}.
However, even this set is {\it not} good for $\xi$-integrability.
Fortunately, transformation to the
both-$\xi$-and-$t$ integrability preserving basis (\ref{Cxi})
from the basis of free cumulants is linear and elementary being given by
the Newton binomial formulas.

In fact, eq.(\ref{dr}) is equivalent to \cite{MMN1}
\be
\sum_{\Delta,R}d_R\varphi_R(\Delta)\chi_R(t)p'(\Delta)=e^{t_1}
\sum_{\Delta,R}d_R\varphi_R(\Delta)\chi_R(t)p'(\Delta)\delta_{|R|,|\Delta|}
\ee
which also implies (\ref{2}). The difference between (\ref{Frof}) and (\ref{Frofg}) for
$k>1$ arises because of the contribution of the structure constants
$C^\Delta_{\Delta_1,\Delta_2}\ne 0$ for $|\Delta|\ne|\Delta_1|$ even if
$|\Delta_1|=|\Delta_2|$.

The algebra of the cut-and-join operators is the Hurwitz theory part of the story.
Many puzzles remain there, including matrix model realizations
\cite{Humamo}, mysterious form of the Virasoro constraints
\cite{Huvira} and an open string generalization to
non-commutative algebra \cite{MMNopen}.
As to the integrability theory part,
it includes relation to the character calculus,
and determinant representations of KP $\tau$-functions.
It summarizes many old developments,
from the studies of Kontsevich matrix models in \cite{GKM}
to those of unitary models in \cite{Unint}.
The story of $\xi$-integrability and the difference between
various choices of $\gamma$ in (\ref{Cxi})
is intimately related to the theory of equivalent hierarchies
\cite{equiv}, a rather sophisticated and not enough widely-known,
though important, chapter of integrability theory.

\section*{Acknowledgements}

Our work is partly supported by Ministry of Education and Science of
the Russian Federation under contract 14.740.11.0081, by NSh 8462.2010.1, by RFBR
grants 10-02-00509-a (A.A., A.Mir. \& S.N.) and 10-02-00499 (A.Mor.),
ANR project GranMa "Grandes Matrices Al\'{e}atoires" ANR-08-BLAN-0311-01 (A.A),
by joint grants 11-02-90453-Ukr, 09-02-93105-CNRSL, 09-02-91005-ANF,
10-02-92109-Yaf-a, 11-01-92612-Royal Society.

\end{document}